\documentclass[11pt]{article}

\usepackage{arxiv}
\usepackage{url}
\usepackage{xcolor}
\usepackage{listings}

\definecolor{mygreen}{rgb}{0.0, 0.8, 0.0}
\definecolor{myred}{rgb}{0.8, 0.0, 0.0}
\usepackage{float}
\floatstyle{plain}
\newfloat{code}{H}{lop}
\floatname{code}{code}
\usepackage{enumitem}
\usepackage[numbers]{natbib}

\usepackage{amsmath,amssymb,amsfonts}
\usepackage{algorithmic}
\usepackage{graphicx}
\DeclareGraphicsExtensions{.png,.pdf,.jpg,.jpeg}
\usepackage{textcomp}

\title{An open-source implementation of a closed-loop electrocorticographic brain--computer interface using Micromed, FieldTrip, and PsychoPy}

\author{
  Bob Van Dyck\\
  Laboratory for Neuro- and Psychophysiology (KU Leuven)\\
  \texttt{bob.vandyck@kuleuven.be}
  \And
  Arne Van Den Kerchove\\
  Laboratory for Neuro- and Psychophysiology (KU Leuven)\\
  \texttt{arne.vandenkerchove@kuleuven.be}
  \And
  Marc M. Van Hulle\\
  Laboratory for Neuro- and Psychophysiology (KU Leuven)\\
  \texttt{marc.vanhulle@kuleuven.be}
}

\date{}

\begin{document}
\maketitle

\begin{abstract}
We present an open-source implementation of a closed-loop Brain-Computer Interface (BCI) system based on electrocorticographic (ECoG) recordings. Our setup integrates FieldTrip for interfacing with a Micromed acquisition system and PsychoPy for implementing experiments. We open-source three custom Python libraries—\textit{psychopylib}, \textit{pymarkerlib}, and \textit{pyfieldtriplib}—each covering different aspects of a closed-loop BCI interface: designing interactive experiments, sending event information, and real-time signal processing.
Our modules facilitate the design and operation of a transparent BCI system, promoting customization and flexibility in BCI research, and lowering the barrier for researchers to translate advances in ECoG decoding into BCI applications.
\end{abstract}

\keywords{Open-source \and Closed-loop \and Real-time \and ECoG \and BCI \and Micromed \and FieldTrip \and PsychoPy}

\section{Introduction}

Brain-Computer Interfaces (BCIs) translate brain activity into useful outputs, enabling various applications including the restoration of communication and muscle control lost to injury or disease~\cite{wolpaw2012, wolpaw2020}. While electroencephalography (EEG) is widely used as a non-invasive method for recording brain activity due to its ease of use and relatively low cost, electrocorticography (ECoG) shows significant potential as a recording technique for permanent medical interventions because of its high spatial and temporal resolution, resistance to noise, and long-term signal stability~\cite{schalk2011, miller2020}. 

Although the validation of ECoG-based BCIs ultimately requires clinical trials involving chronically implanted and optimally located electrodes, acute implants used in presurgical epilepsy monitoring offer a valuable research opportunity. In these clinical settings, the primary objective remains medical but researchers can take advantage of the temporary access to high-quality cortical recordings. However, these settings introduce unique practical challenges that are often underrepresented in the literature. For example, hardware such as the acquisition device—in our case, a Micromed (Italy) system routinely used for epilepsy monitoring—is typically predetermined and not originally intended for real-time applications. Yet, to the authors' knowledge, no published works provide a comprehensive overview of a system architecture addressing the hardware and software challenges under these conditions that could serve as a guide for implementing closed-loop ECoG BCI experiments. To address this gap, we provide a thorough technical description of our system setup alongside the open-source code, aiming to lower the barrier for BCI research in clinical environments.

\begin{figure}
\centerline{\includegraphics[width=0.85\columnwidth]{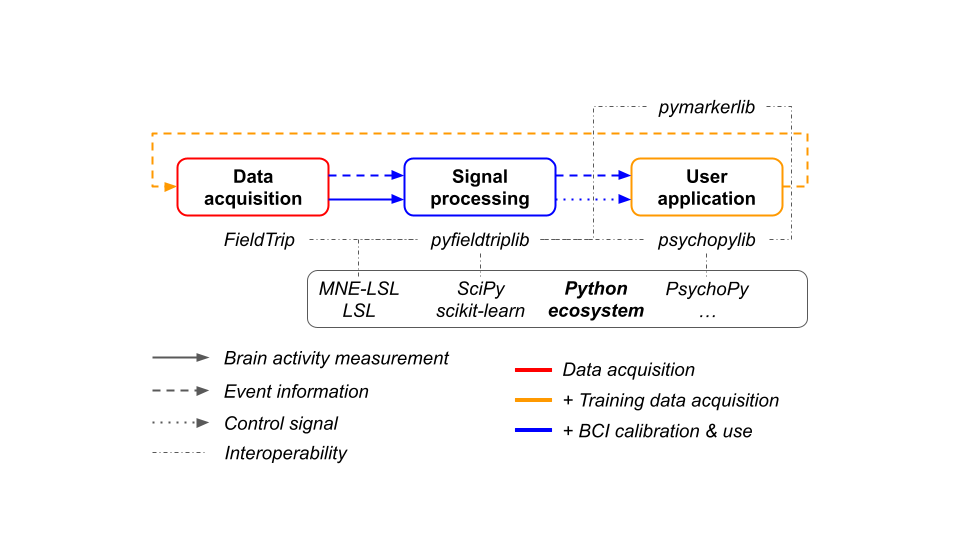}}
\caption{Overview of the three BCI components (data acquisition, signal processing, and user application) and the three stages of design and operation (training data acquisition, BCI calibration, and BCI use).}
\label{fig:bci}
\end{figure}

Conceptually, both EEG- and ECoG-based BCIs comprise three core components—data acquisition, signal processing (segmentation, preprocessing, and modeling), and a user application—and are typically designed and operated in three stages: training data acquisition, BCI calibration, and BCI use \cite{openvibe}.
Fig.~\ref{fig:bci} illustrates the relationship between these components and stages. During training data acquisition, a user application presents stimuli and instructions and sends event information (markers) that provides precise timing and labels, while the data acquisition component continuously records brain activity. In effect, this stage constitutes a conventional experimental paradigm, with stimulus presentation and instructions.
BCI calibration then uses these events to perform event-locked segmentation of the continuous stream into labeled epochs (training data), followed by preprocessing (e.g., resampling, rereferencing, filtering, baseline correction) and model training and selection. In many studies, these processing steps are performed entirely offline (after the session and without planned BCI use) to address research questions. 
Finally, during BCI use, the trained model operates on incoming epochs in real time and its predictions are translated into control commands for the user interface, with the segmentation into epochs performed either synchronously (event-locked) or asynchronously (e.g., using a sliding window). 
A synchronous BCI thus operates at fixed times defined by the system, whereas an asynchronous (self-paced) BCI runs continuously and therefore must also distinguish intentional control from idle periods. 
When the user application delivers feedback to the user contingent on the model predictions, the system constitutes a closed-loop BCI. In research settings, we also refer to this user application as an interactive experiment. 
Note that in a clinical setting, an additional operational stage exists, namely clinical data acquisition or monitoring. 

Because of the different components described above, BCI design requires multidisciplinary expertise in neurophysiology, signal processing, interaction design, computer graphics, and programming. In addition, transitioning from offline to closed-loop BCI experiments often requires bridging a gap in systems engineering expertise. 
A range of free software platforms has been developed to meet these challenges, including BCI2000~\cite{bci2000}, OpenViBE~\cite{openvibe}, Timeflux~\cite{timeflux}, BCI++~\cite{bci++}, and BCILAB~\cite{bcilab}. Among these, only BCI2000 and OpenViBE support direct interfacing with Micromed.
Although these platforms provide feature-rich environments for signal acquisition, real-time processing, and closed-loop feedback, and are both customizable and extensible, fully utilizing them often requires substantial programming knowledge. At the same time, their graphical user interfaces (GUIs), while designed for usability, can restrict workflow flexibility and limit access to lower-level implementation details.
A notable alternative is FieldTrip~\cite{fieldtrip}, a MATLAB toolbox designed primarily for offline and real-time processing of neurophysiological data, which also interfaces with Micromed. While not a full-fledged BCI development platform—lacking integrated stimulus presentation or direct interaction mechanisms, it offers essential tools for signal processing and analysis.
This reflects an inherent trade-off: dedicated BCI platforms provide high-level features such as GUIs but can constrain research flexibility, whereas toolboxes like FieldTrip and EEGLAB~\cite{delorme2004eeglab} offer greater transparency and scripting freedom but lack integrated interfaces for stimulus presentation and user interaction. 
Consequently, researchers often turn to a modular approach, combining multiple specialized tools to construct a flexible and extensible BCI system.

Python libraries are central to the proposed modular approach, as they offer dedicated tools for behavioral experiment design (PsychoPy~\cite{psychopy}), data collection and synchronization (\textit{pylsl}, a Python interface for LabStreamingLayer~\cite{labstreaminglayer}), data analysis (MNE-Python~\cite{mne-python}), and real-time signal processing (MNE-LSL). Their modularity allows researchers to mix and match components for enhanced flexibility and customizability. Embracing a pythonic, modular design philosophy, we can enhance the existing ecosystem by addressing missing functionalities necessary to support a fully functioning BCI system. 


To this end, we introduce the following modules:
\begin{itemize} 
    \item \textit{psychopylib}\footnote{\url{https://gitlab.kuleuven.be/u0140582/psychopylib}}, for implementing a user application in PsychoPy with improved code readability;
    \item \textit{pymarkerlib}\footnote{\url{https://gitlab.kuleuven.be/u0140582/pymarkerlib}}, for sending event information (precise timing and labels) and control signals to data acquisition and other external devices;
    \item \textit{pyfieldtriplib}\footnote{\url{https://gitlab.kuleuven.be/u0140582/pyfieldtriplib}}, for thread-based signal processing that enables the concurrent execution of chained signal processing steps (receiving data streams via LSL or the FieldTrip buffer, segmentation, preprocessing, and modeling) and a user application.
\end{itemize}

Our end-user, hereafter the BCI developer, is a researcher or engineer with Python experience who implements BCIs as modular, Python-defined code and requires a high degree of customizability. In practice, this includes (i) retaining low-level control over real-time signal processing (\textit{pyfieldtriplib}), (ii) designing complex user applications with precise stimulus timing (\textit{psychopylib}), and (iii) sending event information for synchronization with and control over external devices (\textit{pymarkerlib}). 

This paper makes two contributions. First, we provide a detailed technical description of a closed-loop ECoG-based BCI in a clinical context using a Micromed system, offering a practical guide for similar deployments. Second, we introduce three open-source Python libraries—\textit{psychopylib}, \textit{pymarkerlib}, and \textit{pyfieldtriplib}—that support general BCI development, and demonstrate their integration in practice through compact, runnable examples\footnote{\url{https://gitlab.kuleuven.be/u0140582/realtime-bci}} in \ref{app:use-case} and \ref{app:use-case-motor-imagery}. 


\section{System overview}\label{sec:system-overview}

\begin{figure} 
\centerline{\includegraphics[width=0.85\columnwidth]{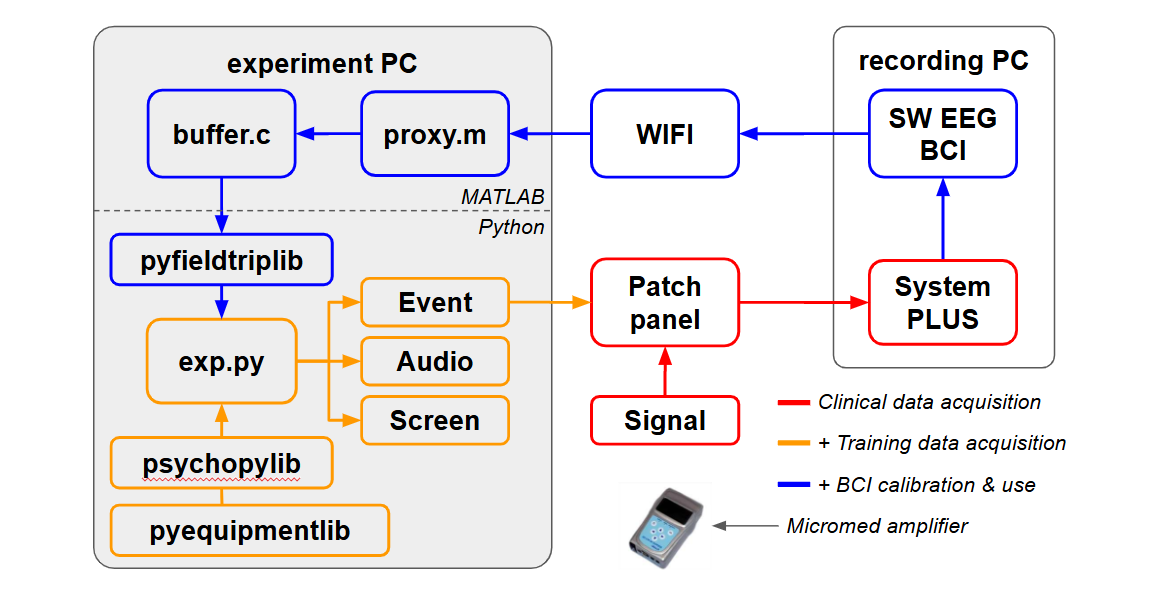}}
\caption{
System overview. 
The setup is divided into three parts corresponding to the three stages of use: clinical data acquisition (orange), training data acquisition (red), BCI calibration and use (blue). 
}
\label{fig:overview}
\end{figure}

Our primary use case is conducting BCI experiments with precise stimulus presentation, real-time neural signal processing, and closed-loop feedback in a clinical setting used for epilepsy monitoring. 
The setup comprises three operational stages, as illustrated in Fig.~\ref{fig:overview}. Data acquisition is managed by a recording PC located in a separate control room, while the signal processing and user application run on a portable experiment PC positioned in front of the participant. Note that the experiment PC is the only component under the BCI developer's direct control. 

\subsection{Clinical data acquisition}

ECoG signals are acquired using an SD LTM 64 Express amplifier (Micromed, Italy) and transmitted via a patch panel (UTP) to a dedicated recording PC in a separate room, where they are stored alongside coregistered recordings (e.g., video footage). 
The SystemPlus EVOLUTION software (Micromed, Italy) processes and saves the data, with the recording PC remotely controlled from the patient’s room through a KVM (Keyboard, Video, and Mouse) switch.
Each recording session generates a TRC file, which stores header information at the start and continuously logs incoming signals throughout the session.

\subsection{Training data acquisition}

During a BCI experiment that only consists of training data acquisition, clinical recording remains unchanged, but additional event information is recorded.
Experiments are implemented in Python using \textit{psychopylib} and executed on the experiment PC. This module leverages PsychoPy for controlling the display, audio, and input collection (e.g., keyboard events, microphone clips).
Event information is transmitted to the recording PC via a standard COM-port serial connection with \textit{pymarkerlib}. 

\subsection{BCI calibration and use}

Real-time signal processing is required for both BCI calibration and BCI use. To this end, recorded signals and event information are transmitted from the recording PC to the experiment PC over a wireless local area network (WLAN).
Data security is ensured by the hospital’s internal protocols. The Transmission Control Protocol/Internet Protocol (TCP/IP) is used for its reliability and robustness, despite its higher latency. 

A recording only starts once a client--server connection is established, which requires an active SW EEG BCI license (Micromed, Italy). This connection ensures that all recorded data packets written to the TRC file are simultaneously transmitted over WLAN, maintaining data integrity. In practice, the SW EEG BCI license can introduce a 30-second gap between consecutive recordings as it searches for a connection before timing out. Since this interruption is unacceptable in clinical settings that require continuous monitoring, we disable the license when no real-time signal processing is needed. 

A FieldTrip proxy, running in MATLAB, acts as an intermediary between the acquisition system and the FieldTrip buffer. The FieldTrip buffer is a lightweight, multithreaded TCP server implemented in C/C++ that stores incoming signals in a ring buffer. It allows data from the acquisition system to be streamed in real-time, while enabling parallel access by multiple clients for processing and feedback. This decoupling ensures that time-consuming computations do not interrupt data acquisition.

The \textit{pyfieldtriplib} module interacts with the FieldTrip buffer and enables real-time signal processing, including epoching and applying user-defined processing chains. 
The resulting processed epochs can be used during a BCI experiment to provide closed-loop feedback. A concrete example of such integration of real-time signal processing and closed-loop feedback in a BCI experiment is provided in \ref{app:use-case}.

\section{Experiment design with psychopylib}

Designing customizable experiments in BCI research, especially those requiring precise stimulus timing and synchronization with neural data, can be complex. PsychoPy~\cite{psychopy} simplifies this by providing a versatile open-source platform with a user-friendly GUI (PsychoPy Builder) for quick design without coding. While the Builder ensures robust and correct execution, the generated code is not meant for direct modification. Using the Builder for increasingly complex experiments can be cumbersome. Although the flexible Code Component feature allows for customization, it relies on programming knowledge, and users with the required expertise may prefer working outside a GUI environment. 

\begin{figure} 
\centerline{\includegraphics[width=0.85\columnwidth]{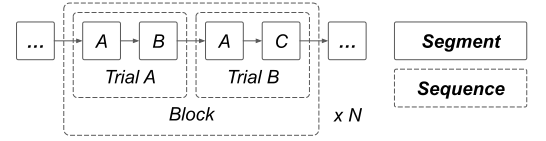}}
\caption{A typical experiment with a block design using \textit{psychopylib} classes. Each trial consists of a sequence of segments in a fixed order, while each block consists of a sequence of trials that are shuffled when the block is repeated (see Listing~\ref{code:psychopylib}).}
\label{fig:psychopylib}
\end{figure}

To address this, the \textit{psychopylib} module provides a more structured, code-based approach to designing BCI experiments. By organizing experiments into sequences of segments, as shown in Fig.~\ref{fig:psychopylib}, it enhances readability and flexibility, making it easier for users with programming experience to implement and customize experiments in PsychoPy.

A \texttt{Segment} represents a quasi-static part of an experiment with a duration and optionally, a key response, a marker value and a set of visual and audio stimuli. When called, it prepares the specified stimuli to be displayed on the next screen update, sends the marker on the screen update, checks for key presses during the segment, and logs once finished. A \texttt{Sequence} represents a sequence of segments that can be repeated and shuffled, has its own (background) visual stimuli, and provides methods to manipulate the sequence, e.g., to set attributes of its segments. When a \texttt{Sequence} object is called, it executes the sequence of segments accordingly. 

\begin{figure}
\centerline{\includegraphics[width=0.85\columnwidth]{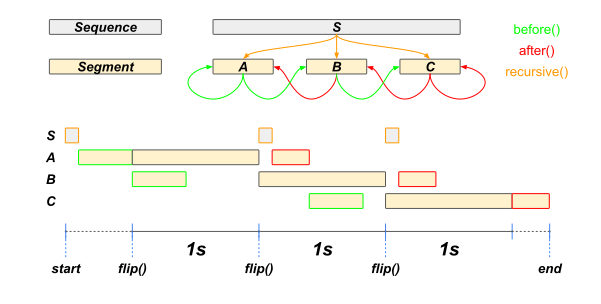}}
\caption{Callback structure when calling a sequence of three one-second segments. Before and after-methods, roughly corresponding to stimulus preparation and logging, are executed during the preceding and subsequent segment, respectively. Sequences also pass along a recursive-method to, for example, draw stimuli associated with the sequence.}
\label{fig:sequence}
\end{figure}

Naively calling one segment after another causes delays due to stimulus preparation and logging tasks before and after each segment. To address this, we employ a callback-based design that ensures precise timing (see Fig.~\ref{fig:sequence}). During the execution of each \texttt{Segment} in a \texttt{Sequence}, the stimuli preparation for the subsequent segment (before-method) and the logging for the preceding one (after-method) are handled. Note that this means that the timing-sensitive portions of an experiment need to be executed as a \texttt{Sequence}. Furthermore, it requires the duration of segments to be long enough for concurrent logging and stimulus preparation, otherwise precise timing is lost.

\begin{code}
\begin{lstlisting}[
language=Python, 
caption={Code for a typical experimental setup with a block design using \textit{psychopylib} classes. Each trial consists of a sequence of segments in a fixed order, while each block consists of a sequence of trials that are shuffled when the block is repeated (see Fig~\ref{fig:psychopylib}).},
label={code:psychopylib}
]
from psychopy.visual import Window, TextStim
from psychopylib.base import Segment, Sequence
# Create a PsychoPy window and stimuli 
win = Window()  
text_a = TextStim(win,'A')
text_b = TextStim(win,'B')
text_c = TextStim(win,'C')
# Create corresponding Segments (of 1 second)
seg_a = Segment(win,1,visual_stim=text_a) 
seg_b = Segment(win,1,visual_stim=text_b)
seg_c = Segment(win,1,visual_stim=text_c)
# Create a sequence of segments (trials)
trial_a = Sequence([seg_a,seg_b])
trial_b = Sequence([seg_a,seg_c])
# Create a sequence of trials (block)
block = Sequence([trial_a,trial_b],shuffle_on_call=True)
# Run the block 5 times and close the window
for _ in range(5): block() 
win.close()
\end{lstlisting}
\end{code}

\section{Event information with pymarkerlib}

The \textit{pymarkerlib} module facilitates the transmission of event information (markers) to equipment used in BCI experiments, ensuring accurate synchronization between the experimental paradigm and recorded neural data. It provides a simple interface for sending event markers to external devices, supporting serial communication protocols among others. The module supports various marker types, including visual markers, serial markers, and specialized markers for the Micromed SystemPlus, as well as devices like eye trackers and VIEWPixx trigger-enabled displays (VPixx Technologies, QC, Canada).

\begin{code}
\begin{lstlisting}[
language=Python,
morekeywords={with,as},
caption={Code for sending event information using \textit{pymarkerlib}. A serial marker with increasing value is sent every second.},
label={code:pymarkerlib}
]
from psychopy.visual import Window, TextStim
from markers.marker import SerialMarker
# Create a PsychoPy window and stimuli 
win = Window() 
refresh_rate = int(1 / win.monitorFramePeriod)
stim = TextStim(win, autoDraw=True)
# Get a marker device 
with SerialMarker(win, port='COM3') as m:
    # Send 10 markers (1 every second)
    for v in range(10):
        stim.text = f"marker: {v:03}"
        m.send(v)
        for _ in range(1 * refresh_rate):
            stim.draw()
            win.flip()
# Close the window
win.close()
\end{lstlisting}
\end{code}
 
\section{Real-time signal processing with pyfieldtriplib}

The \textit{pyfieldtriplib} module employs a multi-threaded architecture to efficiently manage real-time data collection and processing. By allocating separate threads to each processing step and chaining them into a pipeline, the module enables simultaneous data collection and processing without interruptions. This design ensures that the main experimental code remains responsive, avoiding bottlenecks that could otherwise disrupt real-time BCI experiments.

The module provides a clear and flexible framework for data collection. An instance of the \texttt{RtEpoch} class continuously listens for new samples from the FieldTrip buffer and performs epoching. It operates at a configurable rate (default: $60$~Hz), allowing epochs to be defined relative to specific events or using a sliding window approach. These epochs, along with their corresponding event labels, are stored and can be accessed via functions such as \texttt{read\_epoch}. Additional epoching classes accommodate different experimental needs, e.g., \texttt{RtTimeSeries}, a sliding window method for epoching used for asynchronous BCI use.

For data processing, the \texttt{RtFunction} class listens for new epochs and applies user-defined functions to the incoming data. Its event-driven execution ensures efficient resource management while maintaining responsiveness for time-sensitive applications. It also supports MNE-LSL's \texttt{EpochsStream}, enabling the use of alternative signal sources through LSL.

No explicit dejittering, drift correction, or packet reordering is implemented. While TCP/IP can introduce latency and variability in packet timing, event-based epochs retain exact alignment, and any delay impacts only the timing of feedback delivery. The system does not enforce consistency between data availability and feedback presentation; it is up to the user to ensure that feedback is based on the appropriate epoch.

Note that the FieldTrip buffer, implemented as a multithreaded application in C/C++ and compiled into a MATLAB mex file, runs within MATLAB and is accessed via a Python client. As such, it introduces a dependency on MATLAB, which may be a limitation for some users. 

\begin{figure}
\centerline{\includegraphics[width=0.85\columnwidth]{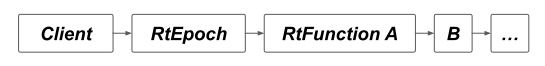}}
\caption{Example of typical experimental setup with a block design using \textit{psychopylib} classes. Each trial consists of a sequence of segments in a fixed order, while each block consists of a sequence of trials that are shuffled when the block is repeated.}
\label{fig:pyfieldtrip}
\end{figure}

\begin{code}
\begin{lstlisting}[language=Python, 
caption={Code for a typical processing pipeline using \textit{pyfieldtriplib}. A client connects to FieldTrip, for each event an epoch is extracted and functions are sequentially applied to each epoch.},
label={code:pyfieldtrip}
]
import time, numpy, scipy.signal
from pyfieldtrip import Client, RtEpoch, RtFunction
# Connect FieldTrip client
client = Client()
client.connect('localhost',5000)
# Epoching with sliding window 
epoching = RtEpoch(client,tmin=0,tmax=0.5)
# Chained processing steps (no_samples,no_channels)
fnc_a = lambda x: scipy.signal.hilbert(x,axis=0)
processing_a = RtFunction(epoching,fnc_a)
processing_b = RtFunction(processing_a,numpy.mean)
# Start threads and collect 50 epochs
threads = [epoching,processing_a,processing_b]
for thread in threads: thread.start()
while processing_b.no_epochs < 50: time.sleep(1)
# Save epochs and features
epoching.save_epochs('./epochs.npz')
processing_b.save_epochs('./features.npz')
# Stop threads and disconnect FieldTrip client
for thread in threads: thread.stop()
client.disconnect()  
\end{lstlisting}
\end{code}

\section{Considerations on latency, decision rate and feedback control}\label{sec:latency}

The system latency of a closed-loop BCI is the delay between the time at which the samples needed for a decision become available at the acquisition system (i.e., the last required sample is digitized) and the time at which the corresponding feedback is realized at the output. Following \cite{wilson2010}, system latency can be decomposed into ADC, processing, and output latency. These stages map to (i) data acquisition and transmission, (ii) segmentation, preprocessing, and model inference, and (iii) feedback delivery. Variability in these delays (latency jitter) is also relevant in practice, as it affects the consistency of feedback timing.

In typical deployments, acquisition-related delays are small compared to the remaining stages. Transmission latency, however, can be substantial and depends on the available connection path (e.g., LAN vs.\ WLAN and network traffic), and is often outside the control of the BCI developer. By contrast, the latency of segmentation, preprocessing, and model inference is defined by the developer and can be reduced through suitable implementation choices. Finally, the feedback delivery stage can be bounded by the update mechanism of the output device (e.g., using PsychoPy for visual feedback, the update is typically realized at the next display refresh).

When implementing user applications with our modules, the BCI developer can account for system latency in several ways. If more predictable feedback timing is desired, estimate an upper bound on the ADC and processing latency and insert a buffer \texttt{Segment} of suitable duration so that feedback is initiated at a fixed delay and the remaining timing variability is due to output latency jitter. If the goal is to minimize latency, structure the application so that processing and feedback are triggered as soon as necessary data becomes available, as in \ref{app:use-case-motor-imagery} where \texttt{Segment} execution is halted until the required epoch is available using the threading Event \texttt{RtEpoch.new\_epoch}.

In asynchronous BCI use, the decision rate refers to how often the system produces a decision (and updates feedback) and can be set independently of system latency. However, if the decision interval is shorter than the system latency, decisions may overlap, increasing queuing delays and CPU load. In practice, the decision rate is tuned to balance responsiveness and computational cost.
With our modules, the decision rate is determined by the epoch arrival rate (set by the hop size in \texttt{RtTimeSeries}) and by how the user application schedules feedback. A simple pattern is to update feedback at a fixed rate matched to the epoch rate, as in \ref{app:use-case}, using a \texttt{LoopingSequence}.

\section{Discussion and conclusion}

In this paper, we present two main contributions toward enabling closed-loop BCI experiments based on ECoG recordings. First, we introduce three modular Python libraries—\textit{psychopylib}, \textit{pymarkerlib}, and \textit{pyfieldtriplib}—which facilitate experiment design, event synchronization, and real-time signal processing, respectively. Together, they provide a flexible toolkit that supports closed-loop BCI implementations without enforcing a rigid framework. Second, we provide a detailed technical description of how these tools can be used in practice, specifically in a clinical setting using a Micromed acquisition system. 

A key aspect of our design philosophy is modularity. While we demonstrated the use of all three modules together, they are not strictly interdependent. For example, \textit{pyfieldtriplib} is completely standalone and can be used independently of \textit{psychopylib}. Users who prefer PsychoPy Builder or alternatives for stimulus presentation can still benefit from our real-time processing and event synchronization tools. Meanwhile, \textit{psychopylib} and \textit{pymarkerlib} build on PsychoPy but do not enforce a rigid structure, allowing users to customize their implementations as needed. This flexibility ensures that researchers can integrate our tools into their existing workflows without unnecessary constraints.

Looking forward, development will focus on translating FieldTrip’s Micromed proxy to Python, eliminating the dependency on MATLAB entirely. Rather than building a large, monolithic system, we continue to emphasize minimalism and modularity—providing lightweight, adaptable components suited to the varied demands of closed-loop BCI research.

\section*{Acknowledgment}
BVD is supported by the Belgian Fund for Scientific Research -- Flanders (G0C1522N) and AVK by the special research fund of the KU Leuven (GPUDL/20/031). MMVH is supported by research grants received from Horizon Europe's Marie Sklodowska-Curie Action (grant agreement No. 101118964). Horizon 2020 research and innovation programme under grant agreement No. 857375, the special research fund of the KU Leuven (C24/18/098), the Belgian Fund for Scientific Research -- Flanders (G0A4118N, G0A4321N, G0C1522N), and the Hercules Foundation (AKUL 043).

\section*{Declaration of generative AI and AI-assisted technologies in the writing process}
During the preparation of this work the authors used Microsoft Copilot and OpenAI GPT-4o/5 in order to improve the readability and language of the manuscript. After using these tools/services, the authors reviewed and edited the content as needed and take full responsibility for the content of the published article.

\appendix

\section{Use case: Closed-loop movement classification}\label{app:use-case}


To demonstrate the integration of the modules introduced above, we implemented a simple ECoG closed-loop movement classifier. Our discussion focuses on the implementation of data streaming, event information, segmentation, real-time processing, and visual feedback, rather than signal processing and BCI performance.

\begin{figure}
\centerline{\includegraphics[width=0.90\columnwidth]{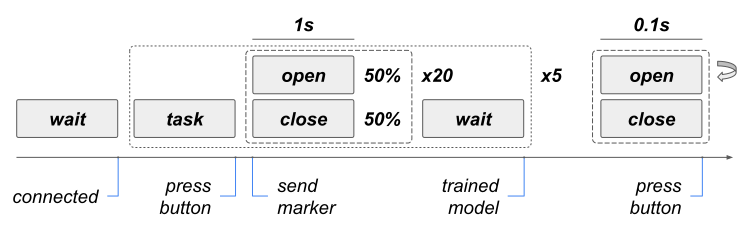}}
\caption{Experimental paradigm (see Listing~\ref{code:paradigm}).}
\label{fig:paradigm}
\end{figure}

\subsection{Experimental paradigm}


The experiment follows our staged framework with training data acquisition, BCI calibration, and BCI use (here, asynchronous).
During training data acquisition, the participant is instructed to open and close their right hand based on a text cue. Each training block consists of $40$ trials, $20$ open and $20$ close, in a randomized order. The open and closed hand positions correspond to rest and to a finger flexion (fist), respectively. When two consecutive trials are the same, the participant sustains the position of the hand. As such, we train a simple state classification model that considers the transition to a state and the maintenance of that state as a single class, making it suitable for a continuous monitoring application. 
In the asynchronous BCI use, the model prediction is updated every $0.1$~s ($10$~Hz decision rate) and shown as text (same as the instruction).

\subsection{Implementation details}

Below, we provide a step-by-step breakdown of the implementation as outlined in Listing~\ref{code:paradigm}, highlighting how the experiment integrates streaming, real-time processing, and modular control using our software stack. The complete example can be found online\footnote{\url{https://gitlab.kuleuven.be/u0140582/realtime-bci}  (paradigms/openclose/)}.

The experiment uses a FieldTrip buffer to stream data either from a live Micromed SystemPlus system over WLAN or from file during debugging. The buffer is initialized in MATLAB, where the IP address of the recording PC must be configured. If not directly known, a tool like Wireshark can be used to discover it. The Python client repeatedly attempts to connect to the FieldTrip buffer, during which a wait screen is shown. This connection loop is managed by a \texttt{ThreadLoopingSequence}, which repeatedly executes a sequence until the thread is terminated.

\begin{lstlisting}[language=Python, label={code:connect}]
client = Client()
thread = ConnectWithRetry(client, 'localhost', 5000)
repeat = ThreadLoopingSequence(thread, [wait])
repeat()
\end{lstlisting}

Signal processing is handled via \textit{pyfieldtriplib}. During training data acquisition, the data is segmented into 1-second epochs ($0$--$1$~s) following event markers 1 (open) and 2 (close). During testing, 1-second epochs are extracted in a sliding window fashion with a 0.1-second hop size (line~\ref{code:p:epoch_test}, Listing~\ref{code:paradigm}). Features from these epochs are used to train a classifier implemented with \textit{scikit-learn}~\cite{scikit-learn}, while the user application is controlled using PsychoPy and \textit{psychopylib}.

The training data acquisition consists of 40 trials (20 open, 20 close), defined using a \texttt{Sequence} object. Trial repetitions can be specified explicitly, or as an additional argument. 

\begin{lstlisting}[language=Python, label={code:train}]
trials_a = Sequence([opened, closed] * 20)
trials_b = Sequence([opened, closed], 20)
assert trials_a.segments == trials_b.segments
assert trials_a.seg_idcs != trials_b.seg_idcs
\end{lstlisting}

The latter method enables advanced shuffle strategies, as it ensures repeated segments have the same index. For instance, we can reshuffle trials until no two adjacent segments are the same.

\begin{lstlisting}[language=Python, label={code:shuffle}]
trials = Sequence([A, B, C], 5)
assert trials.seg_idcs == [0,1,2]
has_no_repeated_elements = lambda idcs: \
    not any(i == j for i, j in zip(idcs, idcs[1:]))
trials.shuffle_until(has_no_repeated_elements)
\end{lstlisting}

Marker devices are instantiated using context managers from \textit{pymarkerlib}. Markers can be set for individual segments (line~\ref{code:p:segment_marker}, Listing~\ref{code:paradigm})) or applied to entire sequences using \texttt{set\_segment\_attr()} (line~\ref{code:p:sequence_marker}, Listing~\ref{code:paradigm})).


A training block then consists of a task instruction, followed by a sequence of trials and a BCI calibration. 
The BCI calibration is implemented as a thread, so the user application remains responsive using \texttt{ThreadLoopingSequence}. 
A limitation of this design is that sequences are tied to the thread in which they were created, and thus cannot be reused after termination of the thread. 


Closed-loop feedback is presented as a visual stimulus determined by the model prediction (control signal). We implement a custom segment. This \texttt{FeedbackSegment} (Listing~\ref{code:paradigm}, line~\ref{code:p:feedback}) reads the latest epoch, performs model inference and sets the feedback stimulus accordingly before proceeding to its execution. 
We repeat this segment indefinitely using a \texttt{LoopingSequence}, which repeats until interrupted by a key press. The segment duration is set to 0.1 seconds to match the epoch hop size, ensuring responsive updates. Finally, by executing this sequence we initiate the asynchronous BCI use, until a termination key is pressed.


\begin{code}
\begin{lstlisting}[language=Python,
caption={Code for training a model from cued instructions, and using the model to provide feedback. The complete example can be found online.},
label={code:paradigm},
numbers={right},
basicstyle={\ttfamily\scriptsize} 
]
# Import necessary libraries and modules
...
# Create a PsychoPy window and stimuli 
win = Window()  
task_stim = TextStim(win,'Open/close your hand')
open_stim = TextStim(win,'Open')
close_stim = TextStim(win,'Close')
wait_stim = TextStim(win,'Wait a moment ...')
# Create corresponding Segments (of 1 second)
task = Segment(win,None,visual_stim=task_stim) 
opened = Segment(win,1,visual_stim=open_stim,marker_value=1)
closed = Segment(win,1,visual_stim=close_stim,marker_value=2)
wait = Segment(win,1,visual_stim=wait_stim)
# Connect FieldTrip client 
... (*\label{code:p:connect}*)
# Epoching cued to events (*\label{code:p:epoch_train}*)
epoching = RtEpoch(client,tmin=0,tmax=1)
epoching.event_ids = [1,2]
epoching.start()
# Processing epochs
def feature_fnc(epoch): ...
processing = RtFunction(epoching,feature_fnc)
processing.start()
# Define sklearn model and training (*\label{code:p:model}*)
model = Pipeline([...])
def fit():
    X, y = processing.read_all_epochs()
    model.fit(X, y)
# Run 5 training blocks of 20 trials (*\label{code:p:train_phase}*)
trials = Sequence([opened,closed], 20, True)
with MicromedSerialMarker(win, port='COM3') as m: 
    task.marker_device = m (*\label{code:p:segment_marker}*)
    trials.set_segment_attr('marker_device', m) (*\label{code:p:sequence_marker}*)
    for _ in range(2):
        task()
        trials()
        # Train model (wait until trained)
        wait_on_fit = ThreadLoopingSequence(
            Thread(target=fit), [wait])
        wait_on_fit()
processing.save_new_epochs('./train.npz')
# Stop epoching 
...
# Restart epoching with sliding window (*\label{code:p:epoch_test}*)
epoching = RtTimeSeries(client,tmin=0,tmax=1, hopsize=0.1)
epoching.start()
processing = RtFunction(epoching,feature_fnc)
processing.start()
# Define feedback sequence (*\label{code:p:feedback}*)
class FeedbackSegment(Segment):
    def _before_segment(self):
        X, _ = processing.read_epoch()
        yp = model.predict([X])
        if yp[0] == 1: feedback_stim = open_stim
        elif yp[0] == 2: feedback_stim = close_stim
        else: feedback_stim = None
        self.vstim = feedback_stim      
        super()._before_segment() 
feedback = LoopingSequence([FeedbackSegment(win, duration=0.1)]) 
# Run real-time feedback (until enter-press)
processing.new_epoch.wait() 
with MicromedSerialMarker(win, port='COM3') as m:
    feedback.set_segment_attr('marker_device', m)
    feedback()
X, y = processing.save_epochs('./test.npz')
# Stop epoching and disconnect
...
\end{lstlisting}
\end{code}

\section{Use case: Closed-loop motor imagery classification from EEG using LabStreamingLayer}\label{app:use-case-motor-imagery}

To demonstrate the correct functioning of the modules, we implemented a canonical motor-imagery BCI paradigm using EEG. 
We report classification accuracy and compare against the literature to demonstrate the correct functioning of our modules. 
We also report end-to-end latency measurements for this BCI system.

\subsection{System overview}

Data acquisition is performed using a Neuroscan SynAmps RT device (Compumedics, Australia), with 32 active Ag/AgCl electrodes distributed over the scalp (QuickCap layout), following the international 10–10 system. The ground electrode was at AFz, and the reference electrode was at POz. When setting up the electrodes, their impedance was kept below 5 $k\Omega$. A recording PC running CURRY 9 (Compumedics Neuroscan) sampled raw EEG at 200~Hz. The software also captured event markers from a VIEWPixx monitor (VPixx Technologies Inc., Canada) and streamed the EEG and event data to LabStreamingLayer (LSL).
The signal processing and user application are executed on an experimental PC connected to the recording PC via LAN. The experimental PC accesses the EEG data through LSL and sends event information to both LSL and the VIEWPixx monitor. We implemented both components in a single Python file that executes training data acquisition, BCI calibration, and BCI use, that can be found online\footnote{\url{https://gitlab.kuleuven.be/u0140582/realtime-bci} (paradigms/motor-imagery/)}. 

\subsection{Experimental paradigm}

During training data acquisition, the participant is instructed to imagine closing their right or left hand based on a text cue. A fixation cross appeared to indicate preparation and when it turned green, participants performed the instructed motor imagery task. Each training block consists of $15$ trials per class, presented in randomized order. 
During BCI use, we maintain the cued execution described above and extend it with post-hoc feedback, presenting a green circle for correct trials and a red circle for incorrect trials.
Each experimental block is followed by a BCI calibration, where the model is re-estimated based on all prior trials.
Note that this continual BCI re-calibration blurs the lines between training data acquisition and BCI use, illustrating the flexibility our modules. 

\subsection{Signal processing}

The signal processing consists of event-based segmentation into epochs (0--2~s relative to the go cue), common average referencing, a filter bank (4~Hz-wide, non-overlapping bands between 8 and 40~Hz), and a common spatial pattern (CSP) filter per band, followed by a linear discriminant analysis (LDA) classifier. Note that the last three components can be jointly described as a filter-bank common spatial pattern (FB-CSP) model, which is widely used in the motor-imagery literature \cite{kai2008}. 


% ...

\subsection{Classification accuracy}

We report classification accuracy for this motor imagery paradigm for a single subject. After the session the subject reported to have experimented with other mental strategies during test block 2 and 3, which explains the drop in performance.

\begin{table}[]
    \centering
    \begin{tabular}{c|c|c|c}
         Test block & 1 & 2 & 3   \\ \hline
         Accuracy (\%)  &  86.67 & 63.33 & 63.33 
    \end{tabular}
    \caption{Classification accuracy during BCI use.}
    \label{tab:placeholder}
\end{table}

\subsection{Latency measurement}

Designing the user application, we opted for an implementation of the post-hoc visual feedback that (i) minimizes latency and (ii) allows measuring the end-to-end latency, by implementing a custom Segment. This \texttt{FeedbackSegment} (Listing~\ref{code:feedbacksegment}) waits until a new epoch is available, performs model inference and sets the feedback stimulus accordingly before proceeding to its execution. Additionally, it sets the Segment's marker value in accordance with the model prediction and sends it when the feedback is presented, allowing us to measure the system latency.

During BCI use, we measured the time difference between go cue and feedback, which consists of the epoch length (2~s) and the system latency (0.178~s $\pm$ 0.016). The relatively high latency is due primarily to non-optimized signal processing, notably the serial execution of the filter-bank’s band-pass filters. This illustrates (i) the dependence of latency on design decisions, and (ii) the fine-grained, low-level control our modules offer the BCI developer. 

\begin{code}
\begin{lstlisting}[language=Python,
caption={Code for implementing a custom Segment (\texttt{FeedbackSegment}) that waits until a new epoch is available, performs model inference and sets the feedback stimulus accordingly before proceeding to its execution.},
label={code:feedbacksegment},
numbers={right},
basicstyle={\ttfamily\scriptsize} 
]
# Import necessary libraries and modules
from psychopy.visual.circle import Circle
from psychopylib.base import Segment
...
# Assume signal processing is implemented
epoching = RtEpochs(...)
def feature_fnc(epoch): ...
processing = RtFunction(epoching, feature_fnc)
model = Pipeline ([...])
# Define custom FeedbackSegment
fb_correct_stim = Circle(win, 50, units='pix', color='green')
fb_incorrect_stim = Circle(win, 50, units='pix', color='red')
class FeedbackSegment(Segment):
    def _before_segment(self):
        # Wait for new epoch
        if processing.new_epoch.wait():
            # Get last epoch and predict
            X, y = processing.read_epoch()
            yp = model.predict(np.array([X]))
            # Set marker value to prediction
            self.marker_value = int(yp)
            # Set feedback stim 
            if bool(yp == y): fb_stim = fb_correct_stim
            else: fb_stim = fb_incorrect_stim            
        else:
            # No new epoch received
            print('No new epoch received')
            fb_stim = None
            self.marker_value = None
            
        # Set stim and continue as usual 
        self.vstim = fb_stim  
        # Proceed as usual 
        super()._before_segment()
        
\end{lstlisting}
\end{code}

\bibliographystyle{ieeetr}
\bibliography{references}

\end{document}